\documentstyle[12pt]{article}
\renewcommand{\theequation}{\arabic{section}.\arabic{equation}}
\newcommand{\eqreset}{\setcounter{equation}{0}}
\newtheorem{theorem}{Theorem}[section]

\newtheorem{lemma}[theorem]{Lemma}
\newtheorem{corollary}[theorem]{Corollary}

\begin{document}
\title{Analytic Bethe ansatz related to a  
 one parameter family of finite dimensional  
 representations of the Lie 
 superalgebra $sl(r+1|s+1)$}
\author{Zengo Tsuboi \\ 
Institute of Physics,                                   
 University of Tokyo,\\ Komaba 
  3-8-1, Meguro-ku, Tokyo 153, Japan}
\date{}
\maketitle
\begin{abstract}
As is well known, the type 1 Lie superalgebra  
 $sl(r+1|s+1)$ admits a one parameter family of 
 finite dimensional irreducible representations. 
 We have carried out an analytic Bethe ansatz related to
  this family of representations. 
We present formulae, which are deformations 
of previously proposed determinant 
 formulae labeled by a Young superdiagram.
 These formulae will provide transfer matrix 
 eigenvalues in dressed vacuum form 
 related to the solutions of a graded Yang-Baxter equation, 
 which depend not only on the spectral parameter but also on a 
 non-additive continuous parameter.   
 A class of transfer matrix functional 
 relations among these formulae is briefly mentioned. 
\end{abstract}
Journal-ref: J. Phys. A: Math. Gen. 31 (1998) 5485-5498 \\ 
DOI: 10.1088/0305-4470/31/24/010 \\\\
Short title: Analytic Bethe ansatz \\ 
\eqreset
\section{Introduction}
 The Analytic Bethe ansatz \cite{R1,R2} is a powerful method 
that postulates the eigenvalues of transfer matrices 
 in solvable lattice models associated with complicated representations  
 of underlying algebras, which are difficult to derive by other method. 
 We can construct systematically them in the dressed vacuum form (DVF) by using  
 Yangians $Y({\cal G})$ \cite{D} analogue of skew-Young tableaux as done in 
 \cite{BR,KS1,KOS} for ${\cal G}=A_{r}$, $B_{r}$, $C_{r}$ and $D_{r}$. 
 
Recently a similar analysis has been done \cite{T2,T3} 
for the Lie superalgebra ${\cal G}=sl(r+1|s+1)$ \cite{Ka} case. 
These results are related to the tensor representations. 
 A class of DVFs are obtained and shown to satisfy a set of functional 
  relations. However, it is well known that 
  the type 1 Lie superalgebras admit a one parameter family of finite 
  dimensional irreducible representations, which is not tensor-like 
  \cite{Ka2}. 
This is also the case with their quantum analogue. 
 Associated with this family of representations, there are solutions 
 \cite{Ma,DGLZ2,DGLZ1} of a graded Yang-Baxter equation, 
 which depend on non-additive continuous parameter. 
  We pointed out \cite{T2} a possibility of extending the DVF 
  related to the tensor representations to the DVF 
 related to a one parameter family of finite dimensional representations.    
 
The purpose of this paper is to extend the DVF \cite{T2} to such 
representations.
  One of the simplest example is $sl(2|1)$ case (cf \cite{RM,PF}): 
 \begin{eqnarray}
 \tilde{{\cal T}}_{1+c}^{2}(u)&=&\frac{Q_{2}(u-1-c)}{Q_{2}(u+1+c)}-
 \psi_{3}(u-1+c)\frac{Q_{1}(u+c)Q_{2}(u-1-c)}{Q_{1}(u+2+c)Q_{2}(u+1+c)}
 \nonumber \\
 &-&
 \psi_{3}(u-1+c)\frac{Q_{1}(u+4+c)Q_{2}(u-1-c)}{Q_{1}(u+2+c)Q_{2}(u+3+c)}
 \label{exam} \\
 &+&\psi_{3}(u+1+c)\psi_{3}(u-1+c)\frac{Q_{2}(u-1-c)}{Q_{2}(u+3+c)}.
 \nonumber	
 \end{eqnarray}
 Note that this function depend on continuous parameter $c$ and 
 still nontrivially pole-free under the Bethe ansatz equation 
(BAE) (\ref{BAE}). 
 We shall construct a large family of the DVF having such features.
  The auxialliary space of 
  the function (\ref{exam}) is related to the finite dimensional 
  representation with the highest weight 
  $(1+c)(\epsilon_{1}+\epsilon_{2})$. For $c\in {\bf Z}_{\ge 0}$, 
  it is tensor representation labeled by the Young superdiagram with shape  
  $((1+c)^{2})$; while for $c\notin {\bf Z}$, it is not
    tensor-like. 
 
  We execute the analytic Bethe ansatz 
based on BAE (\ref{BAE}) associated with the 
distinguished simple root systems of $sl(r+1|s+1)$ \cite{Ka}. 
 Reshetikhin and Wiegmann  observed  \cite{RW} remarkable phenomena 
that the BAE can be expressed 
 by the root system of a Lie algebra
 (see also \cite{Kul} for $sl(r+1|s+1)$ case). 
Furthermore, Kuniba et. al. \cite{KOS} conjectured that the left hand 
side of the BAE (\ref{BAE}) can be expressed
 as a ratio of some \symbol{96}Drinfeld  polynomials' \cite{D}.
 Then one can express the left hand side of the Bethe ansatz 
 equation using the Kac-Dynkin label, which characterizes 
 the quantum space. In view of the fact \cite{Ka2} that one can construct
 a  finite dimensional representation of $sl(r+1|s+1)$
   whose $(r+1)$ th Kac-Dynkin label takes not only  
a non-negative integer value but also a {\it complex} value,
 we assume this is also the case with the 
 LHS of the BAE (\ref{BAE}).
 We introduce the Young superdiagram $\lambda \subset \mu$
 \cite{BB1,DJ} and 
 define the function ${\cal T}_{\lambda \subset \mu}(u)$ 
(\ref{Tge1}), which should be the transfer 
matrix in the DVF whose auxiliary space is a finite dimensional  
tensor module of super Yangian \cite{N,Zh} or 
quantum  affine superalgebra \cite{Y1,Y2},  
labeled by the skew-Young superdiagram $\lambda \subset \mu$; 
while the quantum space is a one parameter family of finite 
dimensional representations which is not tensor-like.
  One can prove the pole-freeness of 
${\cal T}^{a}(u)={\cal T}_{(1^{a})}(u)$ by the same method 
as in \cite{T2}.   
This is also the case with the function
 ${\cal T}_{\lambda \subset \mu}(u)$ since 
 this function has 
determinant expressions whose matrix elements are only 
the functions associated with Young superdiagrams with shape 
$\lambda = \phi $; $\mu =(m)$ or $(1^{a})$. 
  Correspondingly to the complex valued 
 $(r+1)$ th Kac-Dynkin label $b_{r+1}$, 
 we consider a deformation 
$\tilde{\cal T}_{\mu;c}(u)$ of the 
function ${\cal T}_{\mu}(u)$ by a continuous parameter $c$.
This deformation is compatible with the 
so called top term hypothesis \cite{KS1,KOS}. 
We prove the pole freeness of the function  
 $\tilde{\cal T}_{\mu;c}(u)$, 
 an essential property in the analytic Bethe ansatz. 
  Then one may think of the function $\tilde{\cal T}_{\mu;c}(u)$ 
  as a DVF whose auxialliary space and quantum space
   are both parameter dependent. 
 We present a class of transfer matrix functional relations 
among the DVF. It may be viewed as a kind of the $T$-system \cite{KNS1} 
(see also \cite{BR,KOS,T2,T3,Ma,PF,KLWZ,K,KP,KS2,S2,KNS2,KNH,TK,T1}). 

The outline of this paper is given as follows.
In section 2, we briefly review the Lie superalgebra 
${\cal G}=sl(r+1|s+1)$.
In section 3, we execute the analytic Bethe ansatz 
based upon the BAE (\ref{BAE}) associated 
with distinguished simple root systems. 
We note that if we replace the function $\psi_{a}(u)$ with 
the one labeled by the Young superdiagram with shape 
$(1^{1})$, we can reproduce many of our 
earlier results \cite{T2} for 
the function ${\cal T}_{\lambda \subset \mu}(u)$. 
We prove pole-freeness of the function 
$\tilde{\cal T}_{\mu;c}(u)$. 
We briefly mention functional relations
 for the DVF defined in this section. 
 Our main results are the relation (\ref{red1}) and Theorem \ref{main}. 
Section 4 is devoted to summary and discussion. 
Appendix A provides an example of the BAE 
for $sl(2|1)$ with the grading $p(1)=1,p(2)=p(3)=0$. 
Appendix B gives an example of the DVF for $sl(1|2)$. 
\eqreset
\section{The Lie superalgebra $sl(r+1|s+1)$}
In this section, we briefly review the 
Lie superalgebra ${\cal G}=sl(r+1|s+1)$. 
A Lie superalgebra \cite{Ka} is a ${\bf Z}_2$ graded algebra 
${\cal G} ={\cal G}_{\bar{0}} \oplus {\cal G}_{\bar{1}}$ 
with a product $[\; , \; ]$, whose homogeneous
elements obey the graded Jacobi identity.

There are several choices of simple root systems depending on 
 the choices of Borel subalgebras.
The simplest system of simple roots is the so called
 distinguished one \cite{Ka}.
For example,  
the distinguished simple root system
 $\{\alpha_1,\dots,\alpha_{r+s+1} \}$
 of $sl(r+1|s+1)$ has the following form 
 \begin{eqnarray}
   &&\alpha_i = \epsilon_{i}-\epsilon_{i+1}, 
    \qquad i=1,2,\dots,r, \nonumber \\
   &&\alpha_{r+1} = \epsilon_{r+1}-\delta_{1}  \\ 
   && \alpha_{j+r+1} = \delta_{j}-\delta_{j+1} ,
    \quad j=1,2,\dots,s, \nonumber   
 \end{eqnarray}
where  
 $\epsilon_{1},\dots,\epsilon_{r+1};\delta_{1},\dots,\delta_{s+1}$ 
are the basis of the dual space of the Cartan subalgebra with the bilinear 
form $(\ |\ )$ such that 
\begin{equation}
 (\epsilon_{i}|\epsilon_{j})=\delta_{i\, j}, \quad 
 (\epsilon_{i}|\delta_{j})=(\delta_{i}|\epsilon_{j})=0 , \quad 
 (\delta_{i}|\delta_{j})=-\delta_{i\, j}  
\end{equation}
with an additional constraint: 
 \begin{equation}
 \epsilon_{1}+\epsilon_{2}+\cdots +\epsilon_{r+1}-
 \delta_{1}-\delta_{2}-\cdots -\delta_{s+1}=0.
 \end{equation}
 $\{\alpha_i \}_{i \ne r+1}$ are even roots and $\alpha_{r+1}$ 
is an odd root with $(\alpha_{r+1} | \alpha_{r+1})=0$.

Any weight can be expressed in the following form:
\begin{equation}
  \Lambda=\sum_{i=1}^{r+1}\Lambda_{i} \epsilon_{i}
         +\sum_{j=1}^{s+1}\bar{\Lambda}_{j} \delta_{j},
\quad \Lambda_{i},\bar{\Lambda_{j}}\in {\bf C}.
\end{equation}
Let $\lambda \subset \mu$ be a skew-Young superdiagram labeled by 
the sequences of non-negative integers 
$\lambda =(\lambda_{1},\lambda_{2},\dots)$ and 
$\mu =(\mu_{1},\mu_{2},\dots)$ such that
$\mu_{i} \ge \lambda_{i}: i=1,2,\dots;$  
$\lambda_{1} \ge \lambda_{2} \ge \dots \ge 0$;  
$\mu_{1} \ge \mu_{2} \ge \dots \ge 0$ and 
$\lambda^{\prime}=(\lambda_{1}^{\prime},
\lambda_{2}^{\prime},\dots)$ 
be the conjugate of $\lambda $.  
There are two kinds of irreducible tensor representations for $sl(r+1|s+1)$. 
One of them is characterized by the Young superdiagram $\mu$: 
\begin{eqnarray}
&& \Lambda_{i}=\mu_{i} \qquad {\rm for} \quad  
  1 \le i \le r+1 \nonumber \\
&& \bar{\Lambda_{j}}=\eta_{j} \qquad {\rm for} \quad   
 1 \le j \le s+1,
\end{eqnarray}
where $\eta_{j}={\rm max}\{\mu_{j}^{\prime}-r-1,0\}$; 
$\mu_{r+2}\le s+1$.
In this case, the Kac-Dynkin label of $\Lambda$ is 
given \cite{BMR} as follows: 
\begin{eqnarray}
&& b_{j}=\mu_{j}-\mu_{j+1} \quad {\rm for} \quad  
  1 \le j \le r \nonumber\\
&& b_{r+1}=\mu_{r+1}+\eta_{1}  \label{kacdynkin}    \\
&& b_{j+r+1}=\eta_{j}-\eta_{j+1} \quad {\rm for} \quad  
 1 \le j \le s. \nonumber
\end{eqnarray}
A classification theorem for the finite dimensional irreducible 
unitary representations of $gl(r+1|s+1)$ was discussed in 
\cite{GZ}. 
\begin{theorem}\label{class} 
Let $\Lambda$ be a real dominant weight. 
 The irreducible $gl(r+1|s+1)$ module $V(\Lambda)$ with the highest weight 
 $\Lambda$ is  \\ 
 {\rm (1)} typical and type 1 unitary if  
\begin{center} $(\Lambda+\rho,\epsilon_{r+1}-\delta_{s+1})>0 $, 
\end{center}  
 {\rm (2)} or atypical and type 1 unitary if there exists
  $1\le j\le s+1$ such that  
 \begin{center} $(\Lambda+\rho,\epsilon_{r+1}-\delta_{j})=
 0 $. \end{center}
\end{theorem}
 Here $\rho $ is the graded half sum of positive roots: 
 \begin{equation}
 \rho=\frac{1}{2}\sum_{i=1}^{r+1}(r-s-2i+1)\epsilon_{i}+
      \frac{1}{2}\sum_{j=1}^{s+1}(r+s-2j+3)\delta_{j}. 
 \end{equation}
This theorem was generalized to the type 1 quantum 
superalgebra $U_{q}(gl(r+1|s+1))$ for $q>0$ \cite{GS}. 
As remarked in \cite{DGLZ2}, this theorem will be also valid 
for the type 1 quantum superalgebra $U_{q}(sl(r+1|s+1))$ 
for $q>0$ .
Applying Theorem \ref{class} to the aforementioned irreducible tensor 
representation, one finds that \cite{GZ} $\Lambda$ is 
typical and type 1 unitary  if $\mu_{r+1} \ge s+1$; 
atypical and type 1 unitary  if
 $\mu_{r+1} < s+1$.
 
There is a large class of finite dimensional representations \cite{Ka2}, 
which is not tensor-like . 
For example, for the aforementioned irreducible tensor 
representations with the 
highest weight $\Lambda$, a one parameter family of irreducible 
representations with the highest weight (cf. \cite{DGLZ2,GZ}) 
\begin{eqnarray}
&& \Lambda (c)=\Lambda +c \omega ,\label{cweight} \\ 
&& \omega =\epsilon_{1}+\epsilon_{2}+\cdots +\epsilon_{r+1}
\nonumber 
\end{eqnarray}
 is typical and type 1 unitary if 
 \begin{equation}
 (\Lambda(c)+\rho,\epsilon_{r+1}-\delta_{s+1})
 =\mu_{r+1}+\eta_{s+1}-s+c>0.\label{typical}
 \end{equation} 
Note that the $(r+1)$ th Kac-Dynkin label of $\Lambda (c)$ takes  
non-integer value if the parameter $c$ is  
 non-integral. 
 
 The dimensionality of the typical representation of 
 $sl(r+1|s+1)$ with the highest weight 
$\Lambda $ is given 
 \cite{Ka2} as follows 
 \begin{eqnarray}
  {\rm dim}V(\Lambda)&=&2^{(r+1)(s+1)}
   \prod_{1\le i \le j\le r}
  \frac{b_{i}+b_{i+1}+\cdots +b_{j}+j-i+1}{j-i+1}\nonumber \\
  & \times & \prod_{r+2 \le i\le j \le r+s+1}
  \frac{b_{i}+b_{i+1}+\cdots +b_{j}+j-i+1}{j-i+1}.
  \label{dim}
 \end{eqnarray}
 As for the atypical finite dimensional representation, 
 the dimensionality is smaller than the right hand side of (\ref{dim}).
\eqreset
\section{Analytic Bethe ansatz}
Consider the following type of the Bethe
 ansatz equation.
\begin{eqnarray}
 - \prod_{j=1}^{N}
    \frac{[u_k^{(a)}-w_{j}^{(a)}+\frac{b_{j}^{(a)}}{t_{a}}]}
    {[u_k^{(a)}-w_{j}^{(a)}-\frac{b_{j}^{(a)}}{t_{a}}]}
   &=&(-1)^{{\rm deg}(\alpha_a)} 
    \prod_{b=1}^{r+s+1}\frac{Q_{b}(u_k^{(a)}+(\alpha_a|\alpha_b))}
           {Q_{b}(u_k^{(a)}-(\alpha_a|\alpha_b))}, \label{BAE} \\ 
        Q_{a}(u)&=& \prod_{j=1}^{N_{a}}[u-u_j^{(a)}], 
        \label{Q_a} 
\end{eqnarray}
where $[u]=(q^u-q^{-u})/(q-q^{-1})$; $N_{a} \in {\bf Z }_{\ge 0}$; 
$u, w_{j}^{(a)}\in {\bf C}$; $a,k \in {\bf Z}$ 
($1\le a \le r+s+1$,$\ 1\le k\le N_{a}$);
 $t_{a}=1 (1\le a \le r+1)$; $t_{a}=-1 (r+2\le a \le r+s+1)$; 
  $b_{j}^{(a)} \in {\bf Z}_{\ge 0} (1\le a \le r, r+2\le a \le r+s+1)$;
   $b_{j}^{(r+1)} \in {\bf C}$  
  and  
\begin{eqnarray}
    {\rm deg}(\alpha_a)&=&\left\{
              \begin{array}{@{\,}ll}
                0  & \mbox{for even root} \\ 
                1 & \mbox{for odd root} 
              \end{array}
              \right. \\ 
             &=& \delta_{a,r+1}. \nonumber 
\end{eqnarray}
In the present paper, we suppose that $q$ is generic.
The left hand side of the BAE (\ref{BAE}) 
is connected with the quantum space $W=\bigotimes_{j=1}^{N}W_{j}$. 
We assume $W_{j}$ is 
a finite dimensional module of super Yangian \cite{N,Zh} 
or quantum affine superalgebra \cite{Y1,Y2} 
whose classical counter part is characterized by 
 the highest weight with the Kac-Dynkin label  
$(b_{j}^{(1)},b_{j}^{(2)},\dots,b_{j}^{(r+s+1)}) $.
We can find various kinds of the Bethe ansatz equations, 
which are related to the special cases of 
the BAE (\ref{BAE}) 
in many literatures (for example, \cite{Ma,PF,Kul,BKZ,BEF,HGL2,HGL}; 
see also  \cite{KOS,RM,RW,Sc}). 
We suppose that the origin of the left hand side of the 
BAE (\ref{BAE}) goes back to the ratio of some 
\symbol{96}Drinfeld polynomials' $P_{a}(\zeta)
\> (1 \le a \le r+s+1)$
 labeled by the Young superdiagram 
with shape $\mu$:
\begin{eqnarray}
\hspace{-45pt}&&  P_{a}(\zeta) = \prod_{i=1}^{\mu_{a}-\mu_{a+1}} 
     (\zeta-w+a-2 \mu_{a+1}+\mu_{1}-\mu_{1}^{\prime}-2i+1),\>  
     1\le a \le r ,\label{drin1}\\
\hspace{-45pt}&&   P_{r+1}(\zeta) = \prod_{i=1}^{\mu_{r+1}+\eta_{1}} 
     (\zeta -w+r-2\mu_{r+1}+\mu_{1}-\mu_{1}^{\prime}+2i),   \\ 
\hspace{-45pt}&&  P_{r+d+1}(\zeta) = \prod_{i=1}^{\eta_{d}-\eta_{d+1}} 
     (\zeta -w-d+2\eta_{d+1}+r+\mu_{1}-\mu_{1}^{\prime}+2i),
      \quad 1\le d \le s, 
\label{drin3}
\end{eqnarray}
where $\mu_{r+2} \le s+1$;
 $\prod_{i=1}^{0}(\cdots)=1$; $w\in {\bf C}$. 
  One can easily derive these
  polynomials (\ref{drin1})-(\ref{drin3}) by the empirical procedures mentioned 
  in \cite{KOS}. And then we obtain the following ratio of 
  \symbol{96}Drinfeld polynomial' :  
\begin{eqnarray}
  \frac{P_{a}(u+\frac{1}{t_{a}})}{P_{a}(u-\frac{1}{t_{a}})}
  =\frac{u+\frac{b_{a}}{t_{a}}-w^{(a)}}
        {u-\frac{b_{a}}{t_{a}}-w^{(a)}} \label{ratio}
\end{eqnarray} 
where $w^{(a)} \in {\bf C}$;    
 the parameters $\{b_{a}\}$ denote the Kac-Dynkin label
(\ref{kacdynkin}).   
In deriving the relation (\ref{ratio}), we assume the parameters 
$\{b_{a}\}$ are nonnegative integers.
 However as is well known \cite{Ka2},
 one can construct a finite dimensional module whose highest weight is 
labeled by a Kac-Dynkin label with nonnegative integers
 $\{b_{a}\}_{a \ne r+1}$ and a {\it complex} $b_{r+1}$. 
 And then we assume the parameter $b_{r+1}$ in the 
relation (\ref{ratio}) can take non-integer 
value by \symbol{96}analytic continuation\symbol{39}.  
Finally multiplying a natural $q$-analogue of  
(\ref{ratio}) on each site,
 we obtain the left hand side 
of the BAE (\ref{BAE}).

We define the sets 
\begin{eqnarray}
    &&  J=\{ 1,2,\dots,r+s+2\},\nonumber \\
    &&  J_{+}=\{ 1,2,\dots,r+1\}, \quad 
        J_{-}=\{ r+2,r+3,\dots,r+s+2\}   
  \label{set}
\end{eqnarray}
with the total order 
\begin{eqnarray} 
 1\prec 2 \prec \cdots \prec r+s+2 \label{order}
\end{eqnarray}
and with the grading 
\begin{equation}
      p(a)=\left\{
              \begin{array}{@{\,}ll}
                0  & \mbox{for $a \in J_{+}$}  \\ 
                1 & \mbox{for $a \in J_{-}$ }
                 \quad .
              \end{array}
            \right. \label{grading}
\end{equation}
For $a \in J $, set
\begin{eqnarray}
\hspace{-40pt} && z(a;u)=\psi_{a}(u)
     \frac{Q_{a-1}(u+a+1)Q_{a}(u+a-2)}{Q_{a-1}(u+a-1)Q_{a}(u+a)} 
    \qquad    a \in J_{+},
        \nonumber \\
\hspace{-40pt} && z(a;u)=\psi_{a}(u) \frac{Q_{a-1}(u+2r-a+1)Q_{a}(u+2r-a+4)}
        {Q_{a-1}(u+2r-a+3)Q_{a}(u+2r-a+2)} 
    \quad   a \in J_{-},
    \label{z+}
\end{eqnarray}
where $Q_{0}(u)=Q_{r+s+2}(u)=1$.   
 From now on, we will consider the case where the quantum space 
 $W=\bigotimes_{j=1}^{N}W_{j}$ 
is a tensor-product of the module $W_{j}$ labeled by Kac-Dynkin label 
of the form   
 $b_{j}^{(a)}=b_{j}\delta_{a \ r+1}$ ($1\le a\le r+s+1$).  
 In this case, the vacuum part of the function 
$z(a;u)$ takes the following form: 
\begin{equation}
  \psi_{a}(u)=
   \left\{
    \begin{array}{@{\,}ll}
      1 & \mbox{for } \quad a \in J_{+} \\ 
      \prod_{j=1}^{N} 
      \frac{[u-w_{j}+r+1-b_{j}]}
           {[u-w_{j}+r+1+b_{j}]}
        & \mbox{for } \quad a \in J_{-}
    \end{array} \label{psi}
   \right. .
\end{equation}
The generalization to the case of the more general quantum space will 
be achieved by suitable redefinition of the function $\psi_{a}(u)$, and  
such redefinition will not influence the subsequent argument. 
We note that one can recover a function related to the ones in
 \cite{Kul} 
if one set the parameters  
$w_{j}^{(a)},q$ and $ \{ b_{j}^{(a)}\}$ 
in the BAE (\ref{BAE})
 to $0,1$ and the ones in 
 (\ref{kacdynkin}) respectively.  
In this paper, we often express the function $z(a;u)$ as the box
 $\framebox{a}_{u}$, whose spectral parameter $u$ will often 
 be abbreviated. Under the BAE (\ref{BAE}),
 we have 
\begin{eqnarray}
\hspace{-45pt}&& Res_{u=-d+u_{k}^{(d)}}(z(d;u)+z(d+1;u))=0 
    \qquad 1\le d \le r \label{res1} \\
\hspace{-45pt}&& Res_{u=-r-1+u_{k}^{(r+1)}}(z(r+1;u)-z(r+2;u))=0  
     \label{res2} \\
\hspace{-45pt}&& Res_{u=-2r-2+d+u_{k}^{(d)}}(z(d;u)+z(d+1;u))=0 
     \quad r+2\le d \le r+s+1. \label{res3} 
\end{eqnarray}
On the skew-Young  superdiagram 
$\lambda \subset \mu$, we assign a coordinates 
$(i,j)\in {\bf Z}^{2}$ 
such that the row index $i$ increases as we go downwards and the 
column index $j$ increases as we go from left to right and that 
$(1,1)$ is on the top left corner of $\mu$.
Define an admissible tableau $b$ 
on the skew-Young superdiagram 
$\lambda \subset \mu$ as a set of elements $b(i,j)\in J$ 
 labeled by the coordinates 
$(i,j)$ mentioned above, obeying the following rule 
(admissibility conditions).
\begin{enumerate}
\item 
For any elements of $J_{+}$
\begin{equation}
 b(i,j) \prec b(i+1,j),
\label{adm1}
\end{equation}
\item 
 for any elements of $J_{-}$
\begin{equation}
 b(i,j) \prec b(i,j+1)
\label{adm2}
\end{equation}
\item 
and for any elements of $J$
\begin{equation}   
b(i,j) \preceq b(i,j+1),\quad b(i,j) \preceq b(i+1,j).\label{adm3}
\end{equation}
\end{enumerate}
Let $B(\lambda \subset \mu)$ be 
the set of admissible tableaux 
 on $\lambda \subset \mu$.
For any skew-Young superdiagram $\lambda \subset \mu$, 
define the function ${\cal T}_{\lambda \subset \mu}(u)$ as follows
\begin{equation}
 {\cal T}_{\lambda \subset \mu}(u)=
\sum_{b \in B(\lambda \subset \mu)}
\prod_{(i,j) \in (\lambda \subset \mu)}
(-1)^{p(b(i,j))}
z(b(i,j);u-\mu_{1}+\mu_{1}^{\prime}-2i+2j),	
\label{Tge1}
\end{equation}
where the product is taken over the coordinates $(i,j)$ on
 $\lambda \subset \mu$. 
If we replace the vacuum part $\psi_{a}(u)$ (\ref{psi}) of the 
function ${\cal T}_{(1^{1})}(u)$ with the one labeled 
by the Young superdiagram with shape $(1^{1})$,  
 the function ${\cal T}_{(1^{1})}(u)$ corresponds to the   
 eigenvalue formula of the transfer matrix of the 
Perk-Schultz model \cite{Sc,PS1,PS2}
 (see also \cite{Kul}).  
 In this case,  a special case of 
 the function ${\cal T}_{(1^{1})}(u)$ reduces to 
the eigenvalue formula by the algebraic Bethe ansatz
 (For instance, \cite{FK}: 
$r=1,s=0$ case; \cite{EK}: $r=0,s=1$ case; 
 \cite{EKS1,EKS2}: $r=s=1$ case).

  The following relations should be
 valid \cite{T2}. 
\begin{eqnarray}
\hspace{-40pt} && {\cal T}_{\lambda \subset \mu}(u)
 ={\rm det}_{1 \le i,j \le \mu_{1}}
   ({\cal T}_{1}^{\mu_{i}^{\prime}-\lambda_{j}^{\prime}-i+j}
  (u-\mu_{1}+\mu_{1}^{\prime}-\mu_{i}^{\prime}-\lambda_{j}
  ^{\prime}+i+j-1))	
	\label{Jacobi-Trudi1} \\ 
\hspace{-40pt} &&\hspace{30pt}
={\rm det}_{1 \le i,j \le \mu_{1}^{\prime}}
    ({\cal T}_{\mu_{j}-\lambda_{i}+i-j}^{1}
    (u-\mu_{1}+\mu_{1}^{\prime}+\mu_{j}+\lambda_{i}-i-j+1))	,
	\label{Jacobi-Trudi2} 
\end{eqnarray}
where $ {\cal T}_{m}^{a}(u)={\cal T}_{(m^a)}(u)$. 
These relation will be verified by the same method mentioned in 
\cite{KOS}. 
We remark that 
the formula (\ref{Tge1}) reduces to the (classical)
 supercharacter formula  
if we set 
\begin{eqnarray}
\framebox{a} & \to & \exp(\epsilon_{a})
  \qquad {\rm for} \qquad a \in J_{+}, \nonumber \\ 
\framebox{a} & \to & \exp(\delta_{a-r-1})
  \quad {\rm for} \quad a \in J_{-}.  
\end{eqnarray}
 In this case, the functions (\ref{Jacobi-Trudi1}) and 
(\ref{Jacobi-Trudi2})   
reduce to the Jacobi-Trudi formulae on supersymmetric Schur functions 
\cite{BB1,DJ,PT,Kin}. 

The following Theorem is essential in the analytic Bethe ansatz.
\begin{theorem}\label{polefree} {\rm (}{\rm \cite{T2})}
For any integer $a$, the function ${\cal T}_{1}^{a}(u)$  
is free of poles under the condition that
the BAE {\rm (\ref{BAE})} is valid
\footnote{Hereafter singularities of the vacuum parts 
 of the DVFs, which can be removed by multiplying overall scalar functions
  are out of the question.}.   
\end{theorem}
Applying Theorem \ref{polefree} to  (\ref{Jacobi-Trudi1}),
 one can show that 
${\cal T}_{\lambda \subset \mu}(u)$ 
is free of poles under the BAE (\ref{BAE}). 
\begin{sloppypar}
Thanks to the admissibility conditions (\ref{adm1}-\ref{adm3}), 
for any Young superdiagram $\mu$ 
( $\mu_{r+1} \ge s+1$, $\mu_{1}^{\prime} \ge r+1$)  
 and non-negative integer $c$,
only such tableau $b\in B(\mu+(c^{r+1}))$ as  
$b(i,j)=i$ for $1\le i \le r+1,1\le j \le c$; 
$b(i,j)\in J_{-}$ for $r+2\le i \le \mu_{1}^{\prime},1\le j \le \mu_{i}$ 
 is admissible.
 Then the following relation is valid: 
\begin{eqnarray}
\hspace{-27pt} {\cal T}_{\mu+(c^{r+1})}(u)  
&=&
\frac{Q_{r+1}(u-c+\mu_{1}^{\prime}-\mu_{1}-r-1)}
{Q_{r+1}(u+c+\mu_{1}^{\prime}-\mu_{1}-r-1)}
{\cal T}_{\hat{\mu}}(u+\mu_{1}^{\prime}+c-r-1)
\label{red1} \\ 
\hspace{-27pt} &\times &
{\cal H}_{\nu}
(u-\mu_{1}+\mu_{r+2}-c-r-1)  ,
  \nonumber
\end{eqnarray}
where $\mu +(c^{r+1})=(\mu_{1}+c,\mu_{2}+c,
\dots,\mu_{r+1}+c,\mu_{r+2},\dots,\mu_{\mu_{1}^{\prime}})$, 
$\hat{\mu}=(\mu_{1},\mu_{2},\dots,\mu_{r+1})$, 
$\nu=(\nu_{1},\nu_{2},\dots,\nu_{\mu_{1}^{\prime}-r-1})
=(\mu_{r+2},\mu_{r+3},\dots,\mu_{\mu_{1}^{\prime}})$ and 
${\cal H}_{\nu}(u)$ is the function 
${\cal T}_{\nu}(u)$ whose admissible tableaux $B(\nu)$ are 
 restricted to the 
sets of elements of the set $J_{-}$.  
\end{sloppypar}
As a corollary we have (see, \cite{T2})
\begin{eqnarray}
 {\cal T}_{c+s+1}^{r+1}(u)&=&{\cal T}_{((c+s+1)^{r+1})}(u) \nonumber \\ 
&=&
\frac{Q_{r+1}(u-c-s-1)}{Q_{r+1}(u+c-s-1)}
\times
{\cal T}_{s+1}^{r+1}(u+c).\label{red2}
\end{eqnarray}
In deriving the relations (\ref{red1}) and (\ref{red2}), 
we assume $c\in {\bf Z}_{\ge 0}$. 
However these relations will be also valid for $c \in {\bf C}$ 
by \symbol{96}analytic continuation\symbol{39}. 
 We can easily observe this fact from 
 the right hand side of the relations (\ref{red1}) and (\ref{red2}). 
Denote the right hand side of the relations (\ref{red1}) and (\ref{red2}) 
 by $\tilde{{\cal T}}_{\mu;c}(u)$ 
 \footnote{See Appendix B for an example of
  $\tilde{{\cal T}}_{\mu;c}(u)$.}
 and $\tilde{{\cal T}}_{c+s+1}^{r+1}(u)$, 
respectively for arbitrary $c \in {\bf C}$. 
A crucial 
condition for the function $\tilde{{\cal T}}_{\mu;c}(u)$ 
to be the eigenvalue formula of a transfer matrix is given as follows:
 \begin{theorem}\label{main} For any $c \in {\bf C}$, 
the function $\tilde{{\cal T}}_{\mu;c}(u)$  
is free of poles under the condition that
the BAE {\rm (\ref{BAE})} is valid.   
\end{theorem}
As a corollary, we have 
\begin{corollary}\label{coro}
For any $c \in {\bf C}$, 
the function $\tilde{{\cal T}}_{c+s+1}^{r+1}(u)$  
is free of poles under the condition that
the BAE {\rm (\ref{BAE})} is valid.   
\end{corollary}
For any $c \in {\bf Z}_{\ge 0}$, Theorem \ref{main} and 
Corollary \ref{coro} follow from \cite{T2}, while 
for any $c \in {\bf C}$, they require proofs. 
In proving the Theorem \ref{main}, we use the following lemmas.
\begin{lemma}\label{lemmamain}
The function
\begin{equation}
 \frac{{\cal T}_{\hat{\mu}}(u)}
 {Q_{r+1}(u-\mu_{1})} \label{lemmafun}
\end{equation}
is free of poles under the condition that
the BAE {\rm (\ref{BAE})} is valid.   
\end{lemma} 
Proof. 
Thanks to the \cite{T2},  
the function ${\cal T}_{\hat{\mu}}(u)$
 is free of poles under 
the BAE {\rm (\ref{BAE})}. Then we 
 have only to show that the function (\ref{lemmafun})  
 is free of pole at $u=u_{k}^{(r+1)}+\mu_{1}: k=1,\dots ,N_{r+1}$. 
 We will show that ${\cal T}_{\hat{\mu}}(u)$ is divisible 
 by $Q_{r+1}(u-\mu_{1})$. 
In the set $\{ z(a;u+\xi): a \in J, \xi \in {\bf C} \}$, 
 only $z(r+1;u-r+1-\mu_{1})$ and $z(r+2;u-r+1-\mu_{1})$ 
 have $Q_{r+1}(u-\mu_{1})$ in their numerators. 
 So we have only to show that every term
  in ${\cal T}_{\hat{\mu}}(u)$ contains  
  $z(r+1;u-r+1-\mu_{1})$ or $z(r+2;u-r+1-\mu_{1})$. 
 Then all we have to do is to show that $b(r+1,1)=r+1$ or $r+2$ 
 in (\ref{Tge1}) for $\lambda=\phi, \mu=\hat{\mu}$
 since the argument of  $z(b(i,j);u-\mu_{1}+r+1-2i+2j)$ 
 in (\ref{Tge1}) becomes $u-r+1-\mu_{1}$
  only when it's coordinate is $(i,j)=(r+1,1)$.  
 From the admissibility conditions, we can develop the following argument.
  If $b(r+1,1)\preceq r$ then $b(1,1) \prec 1$ since $b(r+1,1)\in J_{+}$; 
  $b(1,1) \prec b(2,1) \prec \cdots \prec b(r+1,1) \preceq r$. 
  This contradicts the fact $b(1,1) \in J$.
  If $b(r+1,1)\succeq r+3$ then $b(r+1,\mu_{r+1}) \succ r+s+2$ since 
  $b(r+1,1)\in J_{-}$; $r+3 \preceq b(r+1,1) \prec 
  b(r+1,2) \prec \cdots \prec b(r+1,\mu_{r+1})$; $\mu_{r+1}\ge s+1$. 
  This contradicts the fact $b(r+1,\mu_{r+1}) \in J$.
 Thus $b(r+1,1)$ must be $r+1$ or $r+2$. 
 In Ref. \cite{T2}, we did not make use of the factor 
 $Q_{r+1}(u-\mu_{1})$ to prove the fact that   
  ${\cal T}_{\hat{\mu}}(u)$ does not have a color $r+1$ pole 
  under the BAE 
  {\rm (\ref{BAE})}. So division by $Q_{r+1}(u-\mu_{1})$ 
  does not influence the proof of the pole-freeness of  
  ${\cal T}_{\hat{\mu}}(u)$ under the BAE 
  {\rm (\ref{BAE})}. Therefore the function (\ref{lemmafun}) 
  is free of poles under 
the BAE {\rm (\ref{BAE})}.   
 \rule{5pt}{10pt} \\ 
\begin{lemma}
{\rm (1)} The function ${\cal H}_{\nu}(u) $ 
is free of color $b$ {\rm (}$b\in J-\{r+1,r+s+2\}${\rm )}
 poles under the condition that
the BAE {\rm (\ref{BAE})} is valid.  \\  
{\rm (2)} The function
\begin{equation}
Q_{r+1}(u-\nu_{1}+\nu_{1}^{\prime}+r+1)
{\cal H}_{\nu}(u) 
\end{equation}
is free of poles under the condition that
the BAE {\rm (\ref{BAE})} is valid.   
\end{lemma}
Proof. (1) One can verify the following relation in the same way as 
the relation (\ref{Jacobi-Trudi1}). 
\begin{eqnarray}
 {\cal H}_{\nu}(u)={\rm det}_{1 \le i,j \le \nu_{1}}
   ({\cal H}_{1}^{\nu_{i}^{\prime}-i+j}
  (u-\nu_{1}+\nu_{1}^{\prime}-\nu_{i}^{\prime}+i+j-1))	
\end{eqnarray}
where $ {\cal H}_{m}^{a}(u)={\cal H}_{(m^a)}(u)$.
Then we have only to show that the function  ${\cal H}_{1}^{a}(u)$ 
is free of color $b$ {\rm (}$b\in J-\{r+1,r+s+2\}${\rm )}
 poles under the BAE {\rm (\ref{BAE})}.
For simplicity, we assume that the vacuum parts are formally trivial, 
that is, the left hand side of the 
 BAE (\ref{BAE}) is constantly $-1$. 
 The function $z(d;u)=\framebox{$d$}_{u}$ with $d\in J $ has 
the color $b$ pole only for $d=b$ or $b+1$, so we shall trace only 
\framebox{$b$} or \framebox{$b+1$} ($b \in J_{-}-\{ r+s+2\}$).
Denote $S_{k}$ the partial sum of ${\cal H}_{1}^{a}(u)$, which contains 
$k$ boxes among \framebox{$b$} or \framebox{$b+1$}.
 Apparently, $S_{0}$ does not have color $b$ pole.
  Thanks to the relation (\ref{res3}), 
$S_{1}$ does not have color $b$ pole ($b \ne r+1$) 
under the BAE (\ref{BAE}). \\
The case $(k \ge 2)$:
 $S_{k}$ is the summation of 
 the tableaux of the form 
\begin{eqnarray} 
&&f(k,n,\xi,\zeta,u):=
\begin{array}{|c|l}\cline{1-1}
    \xi & \\ \cline{1-1} 
    b   & _v \\ \cline{1-1} 
    \vdots & \\ \cline{1-1} 
    b & _{v-2n+2}\\ \cline{1-1} 
    b+1 & _{v-2n} \\ \cline{1-1} 
   \vdots & \\ \cline{1-1} 
    b+1 & _{v-2k+2}\\ \cline{1-1} 
   \zeta & \\ \cline{1-1}
\end{array}  \nonumber \\
&&=\frac{Q_{b-1}(v+2r+3-b-2n)Q_{b}(v+2r+4-b)}
      {Q_{b-1}(v+2r-b+3)Q_{b}(v+2r+4-b-2n)} \\ 
&&\times  \frac{Q_{b}(v+2r+2-b-2k)Q_{b+1}(v+2r+3-b-2n)}
          {Q_{b}(v+2r+2-b-2n)Q_{b+1}(v+2r+3-b-2k)} X 
,\quad  0 \le n \le k, \nonumber
\label{tableauxk3}
\end{eqnarray}
where \framebox{$\xi$} and \framebox{$\zeta$} are columns with 
total length $a-k$, which do not contain \framebox{$b$} and 
\framebox{$b+1$}; $b \in J_{-}-\{r+s+2 \}$; 
$v=u+h$: $h$ is some shift parameter and 
is independent of $n$; the function $X$ does not have color $b$
 pole and is independent of $n$.
$f(k,n,\xi,\zeta,u)$ has color $b$ poles at
 $u=-h-2r-2+b+2n+u_{p}^{(b)}$ and $u=-h-2r-4+b+2n+u_{p}^{(b)}$
  for $1 \le n \le k-1$; at $u=-h-2r-2+b+u_{p}^{(b)}$ 
for $n=0$ ; at $u=-h-2r-4+b+2k+u_{p}^{(b)}$ for $n=k$. 
Obviously, color $b$ residue at 
$u=-h-2r-2+b+2n+u_{p}^{(b)}$
 in  $f(k,n,\xi,\zeta,u)$ and $f(k,n+1,\xi,\zeta,u)$
 cancel each other under the BAE (\ref{BAE}). 
 Thus, under the BAE
  (\ref{BAE}), $\sum_{n=0}^{k}f(k,n,\xi,\zeta,u)$ 
 is free of color $b$ poles ($b \ne r+1$), so is $S_{k}$. \\ 
(2) Among the boxes $\{  \framebox{$a$}: a\in J_{-}\}$, only the 
box \framebox{$r+2$} has color $r+1$ pole. We shall show that the
 color $r+1$ poles in ${\cal H}_{\nu}(u)$, which 
 originate from the box 
  \framebox{$r+2$} are canceled by
   $Q_{r+1}(u-\nu_{1}+\nu_{1}^{\prime}+r+1)$. 
  Owing to the admissibility conditions, \framebox{$r+2$}  
   appears consecutively only at the points $(1,1),(2,1), \dots, (k,1): 
  k \le \nu_{1}^{\prime}$ in each term of ${\cal H}_{\nu}(u)$.
  Then the contribution of \framebox{$r+2$}
   to the term of ${\cal H}_{\nu}(u)$
   which contains $k$ $\ $ \framebox{$r+2$} is  
\begin{eqnarray}
 && \prod_{j=1}^{k} z(r+2,u-\nu_{1}+\nu_{1}^{\prime}-2j+2) \nonumber \\ 
  &=&\frac{Q_{r+1}(u-\nu_{1}+\nu_{1}^{\prime}+r+1-2k)
         Q_{r+2}(u-\nu_{1}+\nu_{1}^{\prime}+r+2)}
        {Q_{r+1}(u-\nu_{1}+\nu_{1}^{\prime}+r+1)
         Q_{r+2}(u-\nu_{1}+\nu_{1}^{\prime}+r+2-2k)}.
\end{eqnarray}
 Thus, the color $r+1$ poles in ${\cal H}_{\nu}(u)$,
  which originated from \framebox{$r+2$} 
  are canceled by $Q_{r+1}(u-\nu_{1}+\nu_{1}^{\prime}+r+1)$. 
  \rule{5pt}{10pt} \\
The dress part of the function $\tilde{{\cal T}}_{\mu;c}(u)$ 
carries $sl(r+1|s+1)$ weight $\Lambda(c)$ (\ref{cweight}). 
One can observe this fact from the 
\symbol{96}top term\symbol{39} \cite{KS1,KOS} of the function. 
The \symbol{96}top term\symbol{39} is considered to be 
related with the highest weight vector.
We speculate the \symbol{96}top term\symbol{39} 
of the function $\tilde{{\cal T}}_{\mu;c}(u)$ 
for large $|q|$ is proportional to  
 \begin{eqnarray}
&& \frac{Q_{r+1}(u-c+\mu_{1}^{\prime}-\mu_{1}-r-1)}
{Q_{r+1}(u+c+\mu_{1}^{\prime}-\mu_{1}-r-1)}
\times 
 \prod_{i=1}^{r+1} \prod_{j=1}^{\mu_{i}} 
 z(i;u-\mu_{1}+\mu_{1}^{\prime}+c-2i+2j)\nonumber \\ 
 &\times & 
 \prod_{j=1}^{s+1} \prod_{i=1}^{\eta_{j}} 
 z(r+j+1;u-\mu_{1}+\mu_{1}^{\prime}-c-2r-2-2i+2j) \nonumber \\ 
 &\approx& 
 q^{-2(\Lambda (c)|\sum_{a=1}^{r+s+1}N_{a}\alpha_{a})}
 =q^{-2\sum_{i=1}^{r+s+1}N_{i}t_{i}b_{i}-2N_{r+1}t_{r+1}c},
 \end{eqnarray}
 where we omit the vacuum part. 
We may think of this circumstance as a generalization of 
 the top term hypothesis \cite{KS1,KOS} to the case of 
 the non-integral highest weight. 
 We believe that the function $\tilde{{\cal T}}_{\mu;c}(u)$ 
 yields actual spectra of the transfer matrix whose auxiliary space 
 is characterized by the highest weight $\Lambda(c)$ 
 at least as long as the typicality condition (\ref{typical}) 
is satisfied. 
 In fact, special cases of the function
  $\tilde{{\cal T}}_{\mu;c}(u)$ 
 are in agreement with the results:  
 for example, for $sl(2|1); \mu=(2^{1})$ case:\cite{RM,PF} 
 (see also \cite{Ma}). 
For the function $\tilde{{\cal T}}_{\mu;c}(u)$, 
 one will be able to use the $R$ matrix 
which is constructed by tensor product graph method 
\cite{DGLZ2,DGLZ1}.    
  
As for negative integer $c$, much care should be taken because 
atypicality condition  may  hold. 
In this case, the dimensionality of the module $V(\Lambda(c))$ 
is no longer the one given by the formula (\ref{dim}).  
For example, for $sl(2|2)$ case,
 $\tilde{{\cal T}}_{1}^{2}(u)$ has the form 
\begin{equation}
\tilde{{\cal T}}_{1}^{2}(u)={\cal T}_{1}^{2}(u)+
 \psi_{3}(u-3)\psi_{3}(u-1){\cal T}_{2}^{1}(u).	
\end{equation}
In this case, the eigenvalue formula in the DVF 
labeled by the Young superdiagram with shape $(1^{2})$ will be the function 
${\cal T}_{1}^{2}(u)$ rather than the function 
  $\tilde{{\cal T}}_{1}^{2}(u)$. 
  
Now we briefly mention the functional relations among the 
functions introduced in this section.
 Thanks to the Jacobi identity,
the following relation holds. 
\begin{equation}
{\cal T}_{m}^{a}(u-1){\cal T}_{m}^{a}(u+1)=
{\cal T}_{m-1}^{a}(u){\cal T}_{m+1}^{a}(u)+
{\cal T}_{m}^{a-1}(u){\cal T}_{m}^{a+1}(u),
\end{equation}
where $a,m \in {\bf Z}_{\ge 0}$.
This functional relation is a specialization of the Hirota bilinear 
difference equation \cite{H} and it is same as the functional relation
 discussed in \cite{T2}
 except the vacuum part. And other functional relations 
  in \cite{T2} are
  also valid except the vacuum part.
Note however that there are another functional relations, 
which arises from a one parameter family of finite dimensional 
representations. For example,
 $\tilde{{\cal T}}_{\mu;c}(u)$ satisfies 
\begin{equation}
 \tilde{{\cal T}}_{\mu;c}(u-d)\tilde{{\cal T}}_{\mu;c}(u+d)
 =\tilde{{\cal T}}_{\mu;c-d}(u)\tilde{{\cal T}}_{\mu;c+d}(u),
\end{equation}
 where $c,d\in {\bf C}$. 
For $\mu=(m^{r+1}), m\in {\bf Z}_{\ge s+1};c=0;d=1$, 
 this functional relation reduces to 
the one in \cite{T2}. 
\eqreset
\section{Summary and discussion}
In the present paper, we have executed the analytic Bethe ansatz 
 related to a one parameter family of finite dimensional  
 representations of the type 1 Lie superalgebra 
 $sl(r+1|s+1)$ 
 based on the Bethe ansatz equations (\ref{BAE}) with distinguished
 simple root system of $sl(r+1|s+1)$. 
 Eigenvalue formulae of transfer matrices in 
 DVF are proposed for a one parameter
  family of finite dimensional representations. 
  The key is the top term hypothesis 
   and the observation that $(r+1)$ th Kac-Dynkin 
   label can take non-integer value. 
 Pole-freeness of the DVF was shown.
 Functional relations have been given for the DVF. 
  
 We emphasize that our method explained in the present paper 
 is still valid even if such factors like gauge factor, extra sign 
(different from $(-1)^{{\rm deg}(\alpha_{a})}$ in (\ref{BAE})), 
 etc. appear in the BAE (\ref{BAE}) 
as long as such factors do not influence the analytical property of
 the right hand side of the BAE (\ref{BAE}). 
  
 There is a remarkable coincidence \cite{FR,FR2} between 
 the free field realization of the generators 
 of $U_{q}({\cal G}^{(1)})$  associated 
 with the classical simple Lie algebras ${\cal G}$ 
and the eigenvalue formulae \cite{KS1} 
in the analytic Bethe ansatz.
 As for a Lie superalgebra ${\cal G}$ case,
 especially in relation with 
 a one parameter family of finite dimensional 
 irreducible representations, no one has discussed  
 such a relation so far. An extensive study will 
 be desirable.
 
  The Lie superalgebras or their
   quantum analogues are not straightforward 
 generalization of their non-super counterparts.
 They have several inequivalent sets of simple root systems 
depending on the choices of their Borel subalgebras. 
  In view of this fact, we generalized \cite{T3} 
 our result \cite{T2} to any simple root system of $sl(r+1|s+1)$. 
  Then we discussed relations among sets of the Bethe 
 ansatz equations for any simple root systems using 
  the particle-hole transformation \cite{BCFH}.
  We pointed out that the particle-hole transformation
     is related with the reflection with respect to the 
   element of the Weyl supergroup for 
  odd simple root $\alpha $ with $(\alpha | \alpha)=0$.

There is another type 1 superalgebra $osp(2|2n)$, 
which also admits a one parameter 
family of finite dimensional representations (see, \cite{ZG,LG}). 
It will be an interesting problem to extend a similar analysis  
discussed in this paper related to $osp(2|2n)$.

Functional relations among fusion transfer matrices at 
{\it finite} temperatures have been given  recently in
 \cite{JKS}  and 
  these functional relations are transformed into 
 TBA equations without using string hypothesis.
 These TBA equations do not carry continuous parameters, 
 which we discussed in this paper. 
  Whether we can derive TBA equations with continuous parameters 
  from our functional relations is an open problem. 
\\
{\bf Acknowledgment} \\  
The author would like to thank Professor A. Kuniba for  
encouragement and useful comments on the earlier version 
of the manuscript. 
\eqreset
\renewcommand{\theequation}{A.\arabic{equation}}
\section*{Appendix A An example of the BAE for 
$p(1)=1,p(2)=p(3)=0$ grading}
Base on the knowledge presented in \cite{KOS}, 
we will consider the BAE for $sl(2|1)$ with the grading 
$p(1)=1,p(2)=p(3)=0$. 
In this case, the simple roots, the sets (\ref{set}) and the 
functions (\ref{z+}) have the form  
$\alpha_{1}=\delta_{1}-\epsilon_{1},
\alpha_{2}=\epsilon_{1}-\epsilon_{2}$,
 $J_{+}=\{2,3\},J_{-}=\{1\}$ and 
\begin{equation}
\framebox{1}=\psi_{1}(u)\frac{Q_{1}(u+1)}{Q_{1}(u-1)},
\framebox{2}=\psi_{2}(u)\frac{Q_{1}(u+1)Q_{2}(u-2)}{Q_{1}(u-1)Q_{2}(u)},
\framebox{3}=\psi_{3}(u)\frac{Q_{2}(u+2)}{Q_{2}(u)}
\end{equation}
 respectively (see, \cite{T2,T3}). 
  The top term labeled by the Young superdiagram with shape $(1^1)$ is 
  proportional to $\framebox{1}$, then we 
  find that the \symbol{96}Drinfeld polynomial' is  
\begin{equation}
P_{a}(\xi)=
  \left\{
     \begin{array}{@{\,}ll}
       \xi-w  & \mbox{for a=1} \\ 
       1      & \mbox{for a=2} 
     \end{array}
  \right. .
\end{equation}  
For any $b\in {\bf Z}_{\ge 1}$, the top term labeled by the Young 
superdiagram with shape $(b^{2})$ will be 
  proportional to  
\begin{eqnarray}
\begin{array}{|c|c|c|c|}\hline 
1 & 2 & \cdots & 2 \\ \hline
1 & 3 & \cdots & 3  \\ \hline 
\end{array}
=\prod_{j=1}^{b+1}\frac{Q_{1}(u+2j+1-b-2)}{Q_{1}(u+2j-1-b-2)} ,
\end{eqnarray}
where we omit the vacuum part. 
 Then we find that the \symbol{96}Drinfeld polynomial' 
 has the following form 
\begin{eqnarray}
P_{a}(\xi)=
  \left\{
     \begin{array}{@{\,}ll}
       \prod_{j=1}^{b+1}(\xi-w-2j+b+2)  & \mbox{for a=1} \\ 
       1 & \mbox{for a=2} 
     \end{array}
  \right. .
\end{eqnarray}
Following \cite{KOS},
 the BAE whose vacuum part corresponds to the quantum space 
 $W=\bigotimes_{j=1}^{N}W_{j}$ 
  labeled by the Young superdiagrams with shape 
 $(b^{2})$ : $j=1$ and  $(1^{1})$ : $2\le j \le N$
 reads as follows: 
\begin{eqnarray}
&& \frac{
   [u_k^{(1)}-w_{1}-b-1]
      }
      {
   [u_k^{(1)}-w_{1}+b+1]
      }
   \prod_{j=2}^{N}
    \frac{
      [u_k^{(1)}-w_{j}-1]
         }
         {
      [u_k^{(1)}-w_{j}+1]
         }
   = 
    \frac{Q_{2}(u_k^{(1)}-1)}
      {Q_{2}(u_k^{(1)}+1)},  \nonumber \\
 &&-1 = 
    \frac{Q_{1}(u_k^{(2)}-1)Q_{2}(u_k^{(2)}+2)}
      {Q_{1}(u_k^{(2)}+1)Q_{2}(u_k^{(2)}-2)},  \label{BAE2}   
\end{eqnarray}
where the parameters $\{t_{a} \}$ are $t_{1}=-1$ and $t_{2}= 1$.
We assume that the parameter $b$ can take non-integer value by  
 \symbol{96} analytic continuation \symbol{39} as in section 3.
 Note that this BAE is in relation to  
 the one in \cite{BEF}. 
The vacuum part $\psi_{a}(u)$ of the function $\framebox{a}$ is 
determined so as to make the function 
${\cal T}_{1}^{1}(u)=-\framebox{1}+\framebox{2}+\framebox{3}$ to be 
 free of pole under the BAE (\ref{BAE2}). 
 Up to an overall scalar function, we have 
 \begin{eqnarray}
 &&	\psi_{1}(u)  =  1
 	\\
 &&	\psi_{2}(u) =\psi_{3}(u) =  
 \frac{
   [u-w_{1}+b]
      }
      {
   [u-2-w_{1}-b]
      }
   \prod_{j=2}^{N}
    \frac{
      [u-w_{j}]
         }
         {
      [u-w_{j}-2]
         }.
 \end{eqnarray}
 Compare the BAE (\ref{BAE2}) to  
 the one (\ref{BAE}) for $sl(2|1)$ 
 with the grading $p(1)=p(2)=0,p(3)=1$  
 and $b_{1}^{(1)}=0;b_{j}^{(1)}=1:2 \le j \le N;b_{1}^{(2)}=b;
 b_{j}^{(2)}=0:2 \le j \le N$,  
  whose vacuum part also  originates from the quantum space 
 $W=\bigotimes_{j=1}^{N}W_{j}$ 
  labeled by the Young superdiagrams with shape 
 $(b^{2}):j=1;(1^{1}): 2\le j \le N$ and 
 analytic continuation argument.
 Note that this BAE is also in relation to the one in \cite{BEF}.
\eqreset
\renewcommand{\theequation}{B.\arabic{equation}}
\section*{Appendix B An example of the DVF}
In this section, we present an example of the DVF
 $\tilde{{\cal T}}_{\mu;c}(u)$
 and the theorem \ref{main} 
for $sl(1|2)$; $\mu=(2,1)$ ; $b_{j}=b$ (in (\ref{psi})); 
$J_{+}=\{1 \};J_{-}=\{2,3 \}$ case: 
\begin{eqnarray}
\tilde{{\cal T}}_{(2,1);c}(u) 
&=&
\frac{Q_{1}(u-c-1)}{Q_{1}(u+c-1)}
 {\cal T}_{(2)}(u+c+1){\cal H}_{(1)}(u-c-2) 
 \nonumber \\ 
&=& 
\frac{\phi(-1-b-c+u)}{\phi(-1+b-c+u)}
 \left\{
 - 
 \frac{Q_{1}(-3 - c + u)Q_{2}(-c + u)}{Q_{1}(3 + c + u)Q_{2}(-2 - c + u)}
 \right. \nonumber \\ 
&-& \frac{\phi(1 - b + c + u)\phi(3 - b + c + u)
         Q_{1}(-1 - c + u)Q_{2}(-4 - c + u)}
       {\phi(1 + b + c + u)\phi(3 + b + c + u)
          Q_{1}(1 + c + u)Q_{2}(-2 - c + u)}
          \nonumber \\ 
 &-& \frac{\phi(1 - b + c + u)\phi(3 - b + c + u)
            Q_{1}(-3 - c + u)Q_{2}(-c + u)}
          {\phi(1 + b + c + u)\phi(3 + b + c + u)
            Q_{1}(1 + c + u)Q_{2}(-2 - c + u)}
           \nonumber \\ 
&+& \frac{\phi(3 - b + c + u)Q_{1}(-1 - c + u)Q_{2}(-4 - c + u)Q_{2}(c + u)}
      {\phi(3 + b + c + u)Q_{1}(1 + c + u)Q_{2}(-2 - c + u)Q_{2}(2 + c + u)}
       \nonumber \\ 
&+& \frac{\phi(3 - b + c + u)Q_{1}(-3 - c + u)Q_{2}(-c + u)Q_{2}(c + u)}
 {\phi(3 + b + c + u)Q_{1}(1 + c + u)Q_{2}(-2 - c + u)Q_{2}(2 + c + u)} 
   \nonumber \\ 
&+& \frac{\phi(3 - b + c + u)Q_{1}(-1 - c + u)
           Q_{2}(-4 - c + u)Q_{2}(4 + c + u)}
         {\phi(3 + b + c + u)Q_{1}(3 + c + u)
         Q_{2}(-2 - c + u)Q_{2}(2 + c + u)} 
         \nonumber \\ 
&+& \frac{\phi(3 - b + c + u)Q_{1}(-3 - c + u)Q_{2}(-c + u)Q_{2}(4 + c + u)}
 {\phi(3 + b + c + u)Q_{1}(3 + c + u)Q_{2}(-2 - c + u)Q_{2}(2 + c + u)}
 \nonumber \\ 
 &-& 
 \left.
 \frac{Q_{1}(-1 - c + u)Q_{2}(-4 - c + u)}
      {Q_{1}(3 + c + u)Q_{2}(-2 - c + u)}
         \right\},\label{exdvf} 
\end{eqnarray}
where 
\begin{eqnarray}
{\cal T}_{(2)}(u) &=&
\begin{array}{|cc|} \hline 
    1 & 1 \\ \hline 
\end{array}
-
\begin{array}{|cc|}\hline
    1 & 2 \\ \hline 
\end{array}
-
\begin{array}{|cc|}\hline
    1 & 3  \\ \hline
\end{array}
+
\begin{array}{|cc|}\hline
    2 & 3  \\ \hline
\end{array} \nonumber 
\\  
&=&
\frac{Q_{1}(-2 + u)}{Q_{1}(2 + u)}
-\frac{\phi(2 - b + u)Q_{1}(-2 + u)Q_{2}(3 + u)}
 {\phi(2 + b + u)Q_{1}(2 + u)Q_{2}(1 + u)} \nonumber \\
&-& \frac{\phi(2 - b + u)Q_{1}(-2 + u)Q_{2}(-1 + u)}
 {\phi(2 + b + u)Q_{1}(u)Q_{2}(1 + u)} \nonumber \\ 
&+& \frac{\phi(-b + u)\phi(2 - b + u)Q_{1}(-2 + u)}
 {\phi(b + u)\phi(2 + b + u)Q_{1}(u)}, 
\end{eqnarray}
\begin{eqnarray}
{\cal H}_{(1)}(u) &=&
-
\begin{array}{|c|} \hline 
    2  \\ \hline 
\end{array}
-
\begin{array}{|c|}\hline
    3 \\ \hline 
\end{array}
\nonumber 
\\  
&=&
-
\frac{\phi(1 - b + u)}{\phi(1 + b + u)}
\left\{ 
 \frac{Q_{1}(-1 + u)Q_{2}(2 + u)}{Q_{1}(1 + u)Q_{2}(u)} 
+
\frac{Q_{2}(-2 + u)}{Q_{2}(u)} 
\right\},
\end{eqnarray}
\begin{eqnarray}
 \phi(u)=\prod_{j=1}^{N}[u-w_{j}].
\end{eqnarray}
The first term in the right hand side of (\ref{exdvf}) is 
the top term, which is related to the highest weight 
$(2+c) \epsilon_{1}+ \delta_{2}$.
Thanks to Theorem \ref{main}, the DVF (\ref{exdvf}) is free of 
pole under the following BAE: 
\begin{eqnarray}
 && \frac{\phi(u_{k}^{(1)}+b)}{\phi(u_{k}^{(1)}-b)}
 =\frac{Q_{2}(u_{k}^{(1)}+1)}{Q_{2}(u_{k}^{(1)}-1)},
\qquad  1 \le k \le N_{1}, \nonumber \\ 
 && -1=\frac{Q_{1}(u_{k}^{(2)}+1)Q_{2}(u_{k}^{(2)}-2)}
         {Q_{1}(u_{k}^{(2)}-1)Q_{2}(u_{k}^{(2)}+2)},
\qquad  1 \le k \le N_{2}.
\end{eqnarray}
We note the fact that if  the parameter $c$ is 
positive integer, $\tilde{{\cal T}}_{(2,1);c}(u)$ has 
a determinant expression whose matrix elements are 
only the functions labeled by Young superdiagrams with 
one column:  
\begin{eqnarray}
\tilde{{\cal T}}_{(2,1);c}(u) &=& {\cal T}_{(2+c,1)}(u) \nonumber \\ 
 &=& {\rm det }_{1 \le i,j \le c+2}
 ({\cal T}_{1}^{\mu_{i}^{\prime}-i+j}(u-c-\mu_{i}^{\prime}+i+j-1)), 
\end{eqnarray}
where $\mu_{1}^{\prime}=2$; $ \mu_{i}^{\prime}=1 : 
2 \le i \le c+2$; $c \in {\bf Z}_{\ge 0}$. 
               
\end{document}